\documentclass[ reprint,
 amsmath,amssymb,
 aps,
 prl
]{revtex4-2}
\usepackage{graphicx}
\usepackage{dcolumn}
\usepackage{bm}
\usepackage{physics}
\usepackage{xcolor}
\usepackage[colorlinks=true,allcolors=blue]{hyperref}

\newcommand{\onestate}{\ket{1}}
\newcommand{\twostate}{\ket{2}}
\newcommand{\Gj}[6]{ \begin{Bmatrix}
  #1 & #2 & #3 \\
  #4 & #5 & #6 
 \end{Bmatrix}}

\begin{document}


\title{Observation of Magnon-Polarons in the Fermi-Hubbard Model}

\author{Max L. Prichard$^1$, Zengli Ba$^1$, Ivan Morera$^{2}$, Benjamin M. Spar$^{1}$,\\ David A. Huse$^{1}$, Eugene Demler$^{2}$ and Waseem S. Bakr$^{1,\dag}$}

\affiliation{$^1$ Department of Physics, Princeton University, Princeton, New Jersey 08544, USA\\
$^2$ Institute for Theoretical Physics, ETH Z\"urich, CH-8093 Z\"urich, Switzerland
}

\date{\today}


\begin{abstract}
The interplay of magnetic excitations and itinerant charge carriers is a ubiquitous phenomenon in strongly correlated electron systems~\cite{mathur_magnetically_1998,dean_persistence_2013,lu_magnetic_2021,tang_simulation_2020,ciorciaro_kinetic_2023}.  In the vicinity of magnetically ordered phases, strong interactions between itinerant quasiparticles and magnetic excitations can result in the dramatic renormalization of both~\cite{bourges_spin_2000,tranquada_quantum_2004,hayden_structure_2004,le_tacon_intense_2011,damascelli2003,koepsell2019imaging,GreinerDynamics2021}. A key theoretical question is understanding the renormalization of the magnon quasiparticle, a collective spin excitation, upon doping a magnetic insulator~\cite{scalapino_common_2012}. Here, we report the observation of a new type of quasiparticle arising from the dressing of a magnon with the doped holes of a cold atom Fermi-Hubbard system, i.e. a magnon-Fermi-polaron. Utilizing Raman excitation with controlled momentum in a doped, spin-polarized band insulator, we address the spectroscopic properties of the magnon-polaron. In an undoped system with strong interactions, photoexcitation produces magnons, whose properties are accurately described by spin wave theory. We measure the evolution of the photoexcitation spectra as we move away from this limit to produce magnon-polarons due to dressing of the magnons by charge excitations. We observe a shift in the polaron energy with doping that is strongly dependent on the injected momentum, accompanied by a reduction of spectral weight in the probed energy window. We anticipate that the technique introduced here, which is the analog of inelastic neutron scattering~\cite{lovesey1984theory}, will provide atomic quantum simulators access to the dynamics of a wide variety of 
excitations in strongly correlated phases.
\end{abstract}

\maketitle

Magnetic fluctuations are believed to play a crucial role in many strongly correlated electronic systems, including unconventional phases in doped cuprates~\cite{scalapino_d-wave_1986,scalapino_fermi-surface_1987,anderson_resonating_1987,gros_antiferromagnetic_1987,zhang_effective_1988,zhang_renormalised_1988,kotliar_superexchange_1988,schrieffer_spin-bag_1988,monthoux_toward_1991,monthoux_toward_1991,monthoux_weak-coupling_1992,dai_evolution_2001,anderson_physics_2004,anderson_physics_2006,edegger_gutzwillerrvb_2007,dahm_strength_2009,le_tacon_intense_2011,dean_persistence_2013}, infinite-layer nickelates~\cite{lu_magnetic_2021,fowlie_intrinsic_2022}, heavy fermion materials~\cite{miyake_spin-fluctuation-mediated_1986,mathur_magnetically_1998,white_unconventional_2015}, Bernal graphene~\cite{zhou_isospin_2022,de_la_barrera_cascade_2022} and moir{\'e} van der Waals systems~\cite{cao_unconventional_2018,sharpe_emergent_2019,chen_tunable_2020,tang_simulation_2020,ciorciaro_kinetic_2023,tang_evidence_2023,xia_superconductivity_2024}.  A natural question that arises in these systems is how magnetic excitations interact with fermionic quasiparticles. For example, this issue arises in understanding the origin of unconventional superconductivity \cite{scalapino_common_2012,keimer_quantum_2015}, where it is commonly argued that the underlying electron pairing mechanism results from an exchange of paramagnons 
(magnetic excitations indicating proximity to the magnetically ordered state) ~\cite{scalapino_d-wave_1986,scalapino_fermi-surface_1987,monthoux_toward_1991,monthoux_weak-coupling_1992}, analogous to the exchange of phonons in conventional superconductors.  In the latter case one can develop a reliable computational scheme even in the case of strong coupling, because electrons and phonons are distinct degrees of freedom and have different intrinsic time scales, which provides the foundation of the Migdal-Eliashberg formalism \cite{migdal_interaction,Eliashberg_interaction}. On the other hand, magnons are collective excitations of electrons composed of the same fermionic quasiparticles. Hence there is no intrinsic separation of energy scales in the microscopic Hamiltonian for magnons and electrons, making the problem theoretically more challenging. One way magnon-electron interactions have been explored experimentally is by studying the fate of magnons in the regime of strong doping of high-T$\rm_c$ cuprates: if superconductivity in the cuprates is mediated by magnetic fluctuations, one expects to see prominent renormalized magnetic excitations even in the overdoped regime. Indeed, addressing this problem has been the subject of many experimental studies during the last few decades
~\cite{bourges_spin_2000,tranquada_quantum_2004,hayden_structure_2004,wakimoto_disappearance_2007,braicovich_dispersion_2009,braicovich_magnetic_2010,le_tacon_intense_2011,dean_persistence_2013,ament_resonant_2011,chaix_resonant_2018,devereaux_inelastic_2007,fujita_progress_2012}.

\begin{figure*}
    \centering
    \includegraphics[width=\textwidth]{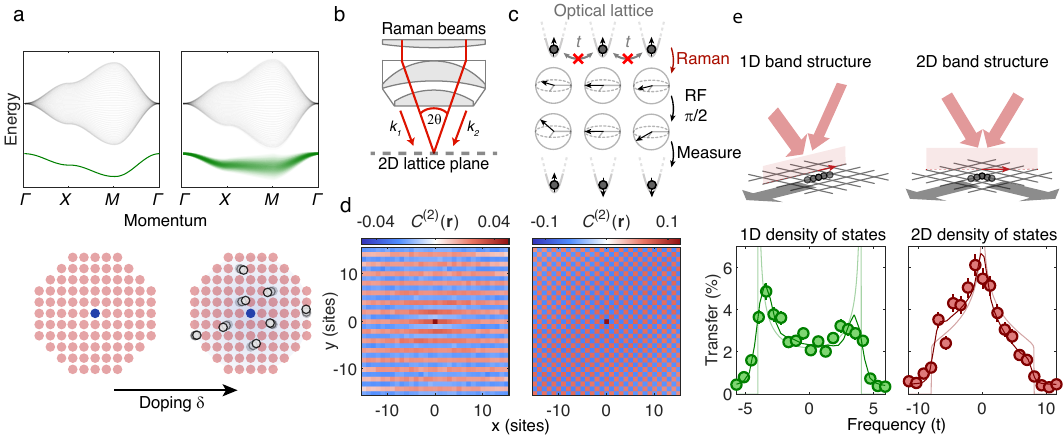}
    \caption{\textbf{Raman spectroscopy of a Fermi-Hubbard system.} (a) Left: schematic spectral function corresponding to a spin-flip injected into a band insulator, showing a well-defined magnon quasiparticle state (green line) and a doublon-hole scattering continuum at higher energies (grey). Right: in the doped system, the spin-flip is dressed by scattered holes, producing a polaron, so the injected magnon has a finite lifetime (green band). (b) The Raman beams are sent through the objective at a controlled angle and illuminate the 2D Fermi-Hubbard system. (c) Injected momentum calibration: application of a Raman pulse followed by a global RF rotation converts the spatially varying Raman phase to a population pattern. (d) Measured two-point spin correlations after the momentum calibration pulse sequence for $\Delta \mathbf{k}\approx (0,1) \, \pi/a$ (left) and $(1,1)\, \pi/a$ (right). The measured momentum is reported from a 2-D sinusoidal fit to the correlations resulting in measured values of $(0.000(1),1.011(1))\,\pi/a$ and $(1.0093(4),0.9941(4)) \, \pi/a$, respectively. (e) Raman spectroscopy of an itinerant, non-interacting Fermi gas at the high symmetry points $\Delta \mathbf{k} \approx (0,\pi)\, a^{-1}$ and $\Delta k \approx (\pi,\pi)\,a^{-1}$. For these momenta, the experimental photoexcitation spectrum approximates the ideal densities of states of a one-dimensional (1D) and two-dimensional (2D) lattice, respectively (lightly shaded curves). Solid lines are numerical simulations taking into account measured temperature, doping and harmonic confinement.}
    \label{fig:Figure1}
\end{figure*}

The unique capabilities of cold atomic systems, particularly the combination of spectroscopic measurements and imaging with single atom resolution, offer a new powerful experimental tool for exploring the interactions of fermionic quasiparticles and collective spin excitations. 
Recent experimental studies have 
examined the renormalization of dopants in an insulating state due to their interaction with collective spin excitations. Specifically, modifications in dopant dynamics \cite{GreinerDynamics2021} and magnetic correlations around dopants \cite{koepsell2019imaging,BakrDirectPolaronImage,GreinerDirectPolaronImage,hartke2023direct}, have been interpreted as signatures of magnetic polarons. However, until now, there have been no experimental studies of how dynamical properties of spin wave excitations, including their dispersion, get modified by finite doping in the Fermi-Hubbard model. In our experiments, we employ Raman spectroscopy to probe the injection of 
spin excitations with non-zero momentum into a hole-doped, spin-polarized band insulator. One of the interesting aspects of our experiments is the continuous tuning both of the interactions and of the doping. 
In the undoped and strongly interacting case, the low-energy states of flipped spins correspond to individual magnons that can be analyzed using the model of spin-exchange interactions (Fig.~\ref{fig:Figure1}a, left). 
With doping, particle-hole excitations of the majority spin species renormalize the spin-flip leading to the formation of a magnon-Fermi-polaron (Fig~\ref{fig:Figure1}a, right). In this setting, the Fermi polaron emerges close to an insulating state which supports magnons as the impurity that gets dressed upon doping. We distinguish this case from the more canonical polaron studied in continuum cold atom experiments~\cite{ZwierleinPolaron,AttrRepl2012,kohstall2012metastabilityNature,cetina2016ultrafast,YanZwierlein2020bose,Ness2020,parish_highly_2013,jorgensen2016,ming2016,YanZwierlein2020bose,darkwah2019observation,duda2023transition}, where the impurity is an atom rather than a collective spin excitation. 


\begin{figure*}[t]
    \centering
    \includegraphics[width=\linewidth]{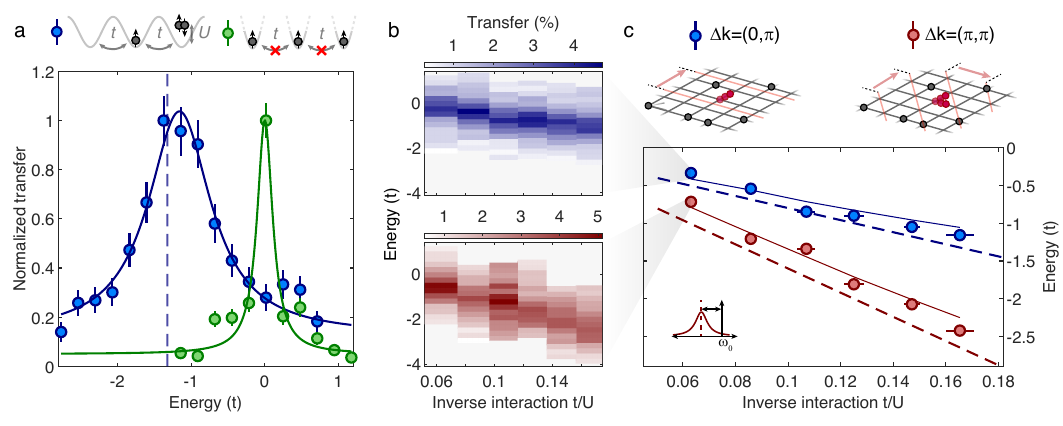}
    \caption{
    \textbf{Interaction dependence of the quasiparticle energy.} (a) 
    Raman spectroscopy 
    with injected momentum $\Delta \mathbf{k}=(0,\pi)a^{-1}$.  Green: Injection spectrum for isolated atoms in a deep lattice. 
    Blue: Injection spectrum for  
    $U/t=6.0(3)$ and low hole doping ($7(2)\%$). Points are transferred atom counts normalized to maximum and solid lines are Lorentzian fits to the data. The frequency axis zero is calibrated to the fitted isolated atom resonance and the dashed line is the energy of a magnon in the Heisenberg model with coupling $J=4t^2/U$. (b) Measured photoexcitation spectra versus inverse interaction $t/U$ at average doping 7(2)\%. The upper (lower) panel corresponds to magnons injected at the $X$ ($M$) point of the BZ. Colorbar displays the density of transferred atoms in the lattice. (c) Magnon 
    energy versus $t/U$. Blue (red) data corresponds to magnon excitation at the $X$ ($M$) point, respectively. Dashed lines are the energy of a single magnon in the Heisenberg model. Solid lines are based on numerics using a non-Gaussian variational wavefunction and include the effects of finite temperature and hole doping.}
    \label{fig:figure2}
\end{figure*}

The starting point of our injection spectroscopy is the creation of a spin-polarized band insulator of $^6$Li in the lowest hyperfine ground state: after evaporating a mixture of atoms in the first ($\ket{1}\equiv\ket{\uparrow}$) and third ($\ket{3}$) lowest hyperfine states to degeneracy, we ramp to the zero-crossing of the $1-3$ scattering length and apply an optical removal pulse to eject all atoms in $\ket{3}$, leaving a thermal sample of atoms in $\ket{\uparrow}$. We then ramp the optical lattice and magnetic field to their final values in 140 ms, depending on the value of the desired final state interactions and tunnel coupling in the lattice. For data taken in the many-body regime, the final lattice depth is $10.6(1) E_R$, where $E_R = h^2/(8 m a^2)=h\times14.7$kHz is the recoil energy for $^6$Li atoms in the lattice with spacing $a = 752 \, \mathrm{nm}$. When there is a second spin state present ($\ket{2}\equiv\ket{\downarrow}$ in this work), the Hamiltonian for $^6$Li atoms in the lattice system can be approximated by the Fermi-Hubbard model $\hat{H}_{\mathrm{FH}}=-t\sum_{\langle i,j\rangle,\sigma}(\hat{c}_{i,\sigma}^{\dagger}\hat{c}_{j,\sigma} + \mathrm{h.c.})+U\sum_i\hat{n}_{i\uparrow}\hat{n}_{i\downarrow}$ where $ \hat{c}^{\dagger}_{i\sigma} $ ($\hat{c}_{i\sigma}$) creates (destroys) a fermion of spin $\sigma$ at lattice site $i$, the number operator is $\hat{n}_{i\sigma} = \hat{c}^{\dagger}_{i\sigma}\hat{c}_{i\sigma}$, 
$\expval{i,j}$ denotes nearest-neighbor sites, $t$ is the nearest-neighbor tunneling rate and $U$ in the on-site interaction.

We then illuminate the sample through a high numerical aperture (NA) objective with two Raman beams coupling $\ket{\downarrow}$ to  $\ket{\uparrow}$ with strength $\Omega_R$. Both the detuning $\Delta$ from the two-photon resonance and the transferred Raman momentum $\Delta \bf{k} = \bf{k_2} - \bf{k_1}$ are tunable in our experiment (Fig.~\ref{fig:Figure1}b). The rotating frame Hamiltonian $H_{\mathrm{R}}=\frac{1}{2}\hbar\Omega_R\sum_j(\hat{c}^\dag_{j,\downarrow}\hat{c}_{j,\uparrow} e^{i \Delta \mathbf{k}\cdot \mathbf{r}_j}+\mathrm{h.c.})+ \frac{1}{2}\hbar \Delta (\hat{n}_{\uparrow,j}-\hat{n}_{\downarrow,j})$ thus describes a spin-flip with a controlled momentum transfer. Upon evolving under the combined Hamiltonian $H =H_{\mathrm{FH}} + H_{\mathrm{R}}$ for a variable amount of time, we freeze evolution of the many body state by simultaneously quenching the lattice depth to approximately $60\, E_R$ and extinguishing the Raman light, after which the transferred atoms are imaged with fluorescence imaging.

Interference between the Raman beams with frequencies $\omega_1$ and $\omega_2$ creates a running wave with electric field $\mathbf{E}(\mathbf{r},t) \sim e^{i (\omega_1-\omega_2) t - \Delta\mathbf{k}\cdot\mathbf{r}}\equiv e^{i \Delta\omega \,t}e^{i\phi(\mathbf{r})}$ which is responsible for the transferred momentum in an extended system. For a set of fully localized and independent two-level systems however, this is equivalent to Larmor precession in a magnetic field with spatially varying projection in the $x-y$ plane. We approximate this scenario using a spin-polarized band insulator in a lattice of depth $33~E_R$. Following evolution under $H_\mathrm{R}$ alone, the spin populations are measured in a basis along the equator of the Bloch sphere using a global $\pi/2$ radiofrequency pulse prior to readout (Fig.~\ref{fig:Figure1}c). The resulting fluoresence images display long-range density correlations (Fig.~\ref{fig:Figure1}d) which are a direct probe of $\phi(\mathbf{r})$, allowing for a precise determination of $\Delta \mathbf{k}$ (Methods). The available NA of our objective allows for continuous tunability of the injected Raman momentum in and beyond the first Brillouin zone (BZ) of the square lattice.

Having benchmarked the transferred momentum with localized atoms, we first turn to probing itinerant non-interacting fermions by tuning the magnetic field to the zero-crossing of the $\ket{1}-\ket{2}$ scattering length at approximately 527 Gauss. For extended single particle Bloch states $\ket{n,\mathbf{q}}$ with band index $n$ and quasimomentum $\mathbf{q}$, the Raman coupling drives transitions between $\ket{\uparrow,n,\mathbf{q}}$ and $\ket{\downarrow,n', \mathbf{q}'}$, where $\mathbf{q}$ and $\mathbf{q}
'$ obey momentum conservation up to that provided by the reciprocal lattice momentum. As a range of initial momenta are occupied by the Fermi sea, the overall transferred spectrum will reflect both the initial filling as well as the final density of states of the system at the chosen momentum.  In particular, the momenta $(0,\pi)a^{-1}$ and $(\pi,\pi)a^{-1}$ correspond to the $X$ and $M$ points of high symmetry in the Brillouin zone, and for the limiting case of a unity filled band insulator, the transfer fraction reduces to the exact density of states for non-interacting particles in a 1D and 2D lattice respectively (Methods). The resulting spectra clearly reflect the dimensionality, with a characteristic suppression (enhancement) of spectral weight at zero frequency for the $X$ ($M$) point (Fig.~\ref{fig:Figure1}e). We fit both spectra independently to the results of a simulation including the effects of finite temperature, non-zero hole doping and overall harmonic confinement, and find good agreement for $t_x = t_y = 433(7)$ Hz (solid lines).


\begin{figure*}[t]
    \centering
    \includegraphics[width=\linewidth]{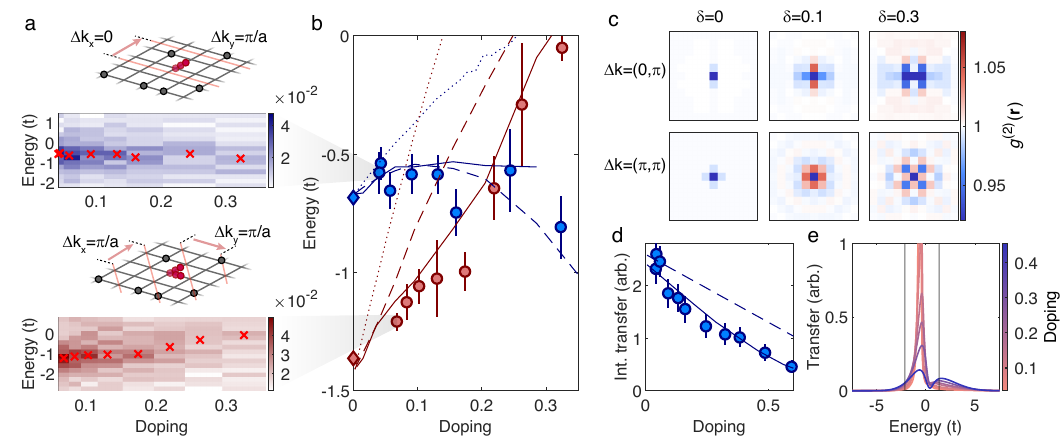}
    \caption{\textbf{Doping dependence of the magnon-polaron energy.} (a) Photoexcitation spectra versus doping taken at the $X$ point (top) and the $M$ point (bottom). For all dopings where the signal is strong enough to fit a peak, we overlay the peak transfer resonance as a guide to the eye. (b) Polaron binding energy at the $X$ point (blue) and $M$ point (red). Experimental data (circles) is compared to three variational calculations: molecular Chevy ansatz (dotted lines), non-Gaussian ansatz ground state (dashed lines) and non-equilibrium non-Gaussian ansatz (solid lines). All theories converge to the magnon energies at zero doping (diamonds). (c) Calculated $g_2$ density correlation function around the impurity in the polaron frame as a function of doping $\delta$ and transferred momentum. (d) Integrated atom transfer within the probed detuning window $\Delta\in[-2.1,1.3]t/h$ as a function of doping at the $X$ point. Experimental transfer (circles) is compared to the theoretical transfer integrated over the experimental window (solid lines) and all calculated frequencies (dashed line) with an overall scaling factor as a free parameter. (e) Theoretical spectra versus energy at the $X$ point, showing spectral weight leaving the experimentally probed window (vertical lines) as a result of dressing with high-energy excitations.}
    \label{fig:VersusDoping}
\end{figure*}

We now turn to analyzing the case of an interacting final state. In the linear response regime, the transfer rate during the photoexcitation pulse is given by the relation
\begin{equation}
    \Gamma_\beta({\textbf{k}},\omega) = \frac{2\pi}{\hbar} |\Omega_R|^2 S^-_\beta({\textbf{k}},\omega),
\end{equation}
where $S_\beta^-(\mathbf{k},\omega)$ is the (finite temperature) dynamical spin structure factor \cite{EugeneIvanSpecTheory}. As a direct probe of $S^-(\mathbf{k},\omega)$, our experimental technique is the analog of inelastic neutron scattering (INS), a critical tool in the study of magnetic ordering and excitations in condensed matter systems \cite{hayden_structure_2004,vaknin1987antiferromagnetism,tranquada_quantum_2004,fong_spin_2000,fong1999neutron}. Photoexcitation spectra can be obtained for different interactions by changing the magnetic field in the vicinity of the Feshbach resonance at 834 Gauss. We determine the quasiparticle energy shifts in the interacting system (which are typically in the sub-kHz range) by comparing the measured resonance in a 10.6(1) $E_R$ lattice with the resonance in a deep $33 E_R$ lattice where tunneling is negligible and the spectra reflect the single-atom resonance. A sample spectrum for this case is displayed in Fig~\ref{fig:figure2}a.

In Fig.~\ref{fig:figure2}b, we plot the measured spin spectral function versus inverse interactions $t/U$ from the regime of intermediate coupling $U/t=6.1(2)$ to the strongly interacting regime $U/t=15.9(6)$. For this analysis, we focus on the central region of the lattice where the peak filling is high (93(2)\%), and the transfer fraction is low ($\sim$5\%), suppressing non-linear effects in the spectroscopy. We find that the spectra at all interactions fit well to a Lorentzian function, allowing us to extract the energy shift from the center frequency. In Fig.~\ref{fig:figure2}c we show the measured energy shifts versus $t/U$ at both $\mathbf{k}\neq 0$ high-symmetry points in the first BZ. For an undoped system and in the regime of strong interactions where double occupancies are suppressed, the Hubbard model can be approximated by an antiferromagnetic Heisenberg model $\mathcal{H}= J \sum_{\langle i,j\rangle} \mathbf{S}_i \cdot \mathbf{S}_j $, where $\mathbf{S}_i$ is the spin on site $i$ and the spin coupling is given by $J=4t^2/U$. We compare the measured energies in this low doping regime to the predictions of the linear spin wave theory of the Heisenberg model (dashed lines), where the elementary excitations are magnons with dispersion $E(\mathbf{k}) = -J(2-\cos(k_xa)-\cos(k_ya))$. While we obtain reasonable agreement, we find that the measured quasiparticle energy shift is generally overestimated by this simple prediction. A better agreement is obtained by accounting for the nonzero temperature and hole density using a more refined theoretical treatment as described below (solid lines). The Loretzian fits also allow us to extract the linewidth of the quasiparticle peaks (Methods), which in all cases is well above the linewidth obtained in the case of isolated atoms (Fig.~\ref{fig:figure2}a). Magnon-hole scattering in the doped system leads to a finite lifetime of the injected magnons, a fundamental linewidth broadening mechanism even at zero temperature. However, our numerics indicate that other broadening mechanisms also play a role in our system, including the overall harmonic confinement as well as the finite temperature of the gas (Methods).

Owing to the confining potential from the lattice beams, a range of densities are present in the initial spin-polarized gas. To probe the evolution of the spectra with doping, we analyze the distribution of transferred atoms after the Raman pulse in a series of radial bins, resulting in a range of hole densities off of which the injected magnon may scatter. The doping dependence of the raw spectra is displayed in Fig.~\ref{fig:VersusDoping}a for $U/t = 11.7(3)$ at both the $M$ and $X$ high-symmetry points in the BZ. From the evolution of the spectral distribution upon doping, it is clear that the character of a single magnon is strongly renormalized in the presence of hole dopants, which we observe as both a frequency shift and a redistribution of spectral weight.  

In Fig.~\ref{fig:VersusDoping}b we show the fitted quasiparticle energy versus doping for both momenta and compare with various theoretical calculations. In the limit of low doping, the data and all the theoretical methods approach the single magnon energy in the Heisenberg model (diamonds). However, as the system is doped away from a band-insulating initial state, the evolution of the magnon-polaron energy is qualitatively different for the two excitation momenta. For the case of excitation at the $X$ point, we find that the resonance remains approximately constant for dopings up to $25\%$, while for the $M$ point the resonance shifts quickly with doping, with the latter energy approaching zero at around $30\%$ doping, despite having an 
energy at zero doping almost twice as low as the $X$ point energy. 

We compare the data to three theories with increasing levels of sophistication. Experimentally measured binding energies of polarons in spin-imbalanced continuum gases have been accurately reproduced using the Chevy ansatz~\cite{ChevyVariation,ZwierleinPolaron}. In analogy with the molecular Chevy ansatz \cite{Mora2009,Punk2009}, we introduce a variational wavefunction $\ket{\phi_{\mathbf{k}}}$ for a polaron with momentum $\bf{k}$,
\begin{equation}
    \ket{\phi_{\mathbf{k}}}
    = \sum_{\mathbf{q}}\phi_{\mathbf{q},\mathbf{k}}\hat{c}^{\dagger}_{\mathbf{k}+\mathbf{q},\downarrow} 
 \hat
    {c}_{\mathbf{q},\uparrow}\ket{\mathrm{FS}}_\uparrow\otimes\ket{0}_\downarrow,
    \label{Eq:Ren_Mag_wavef}
\end{equation}
where $\ket{\mathrm{FS}}_\uparrow$ denotes the majority Fermi sea and $\phi_{\mathbf{q,k}}$ are variational parameters. We note that while $\ket{\phi_\mathbf{k}}$ only includes a single hole excitation of the Fermi sea induced by the Raman beams, this ansatz is actually a solution for a single hole dopant in the band insulator and captures doublon-hole fluctuations at all orders in $t/U$ for vanishing doping, going beyond the superexchange paradigm. The energy obtained using this ansatz (Fig.~\ref{fig:VersusDoping}b, dotted lines) qualitatively captures some features of the data, including the almost linear increase of the energy with doping at the $M$ point, but significantly underestimates the binding energy for finite doping. This points to the need for a more refined theory which includes higher order particle-hole fluctuations. We take these into account using a variational Gaussian state in the frame of reference of the polaron (a non-Gaussian state in the lab frame). This is achieved using a Lee-Low-Pines transformation, which eliminates the impurity degree of freedom at the cost of introducing interactions between the particles of the spin-polarized Fermi sea (Methods). The ground-state energy calculated with this approach already exhibits improved agreement with the data (dashed lines), while extending the approach to a finite-temperature non-equilibrium calculation of the spin spectral function yields excellent agreement with the data (Methods). These results indicate a transition at around $20\%$ doping where the minimum of the magnon-polaron dispersion switches from the $M$ point for low doping to the $X$ point at high doping, a trend which is also present in the experimental data.\

\begin{figure}[t]
    \centering
    \includegraphics[width=\linewidth]{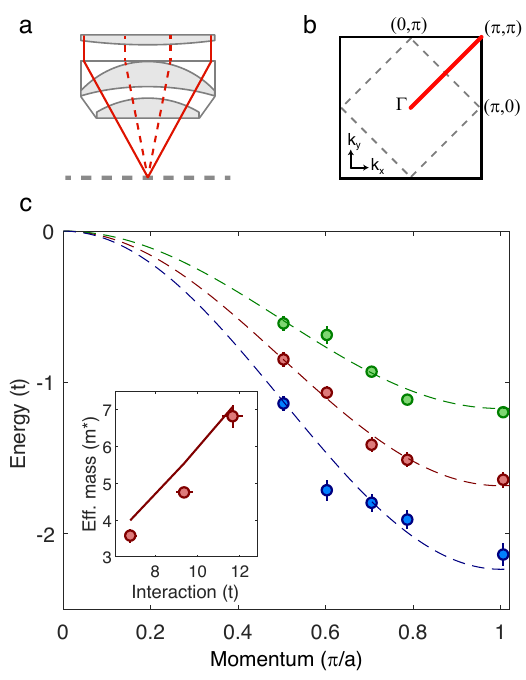}
    \caption{
    \textbf{Magnon-polaron dispersion and effective mass. }(a) Tuning the injected magnon momentum. Adjusting the position of the Raman beams in the Fourier plane of the objective tunes the transferred momentum $\Delta \mathbf{k}$. (b) First Brillouin zone of the square lattice. Momentum scans are taken along a line extending from the $M$ point $(\pi,\pi)$ to the $\Gamma$ point (red line) to characterize the dispersion of the polaron along this line. (c) Measured dispersion of the polaron for $U/t=11.7(5)$ (green), 8.3(5) (red) and 7.0(2) (blue). Dashed lines are fits to a single magnon dispersion with the bandwidth as a fit parameter. Inset: the effective mass extracted from the fitted dispersion (circles) together with the calculated effective mass of the polaron (solid line) in units of $m^*=\hbar^2/{2a^2 t}$, the effective mass of a single atom in an empty square lattice.}
    \label{fig:vsMomentum}
\end{figure}

To gain insight into this behavior, we consider the calculated real-space correlation function $g^{\mathrm{(2)}}(\mathbf{r})\equiv \langle\hat{n}_{\mathbf{r},\uparrow} \hat{n}_{\mathbf{0},\downarrow}\rangle/n_\uparrow n_\downarrow$ between the spin down impurity and the majority spin up atoms in Fig.~\ref{fig:VersusDoping}c (Methods). From these plots we find a significant difference in the spatial structure of the magnon-polaron depending on the injected momentum. For excitation at the $M$ point with $10\%$ doping of the insulator, the injected magnon displays positive correlations for all nearest-neighbor and next-nearest neighbor $\ket{\uparrow}$ fermions. This indicates an effective repulsion to the presence of additional holes and explains the rapid shift towards positive energies observed in Fig.~\ref{fig:VersusDoping}b. In contrast excitations at the $X$ point have markedly anisotropic correlations, indicating an effective magnon-hole repulsion along the direction of momentum transfer and an attractive interaction along the orthogonal direction. Therefore, for these low dopings, the mean-field magnon-hole interaction vanishes, resulting in the markedly suppressed shifts with doping observed in Fig.~\ref{fig:VersusDoping}b.


In addition to the energy shift, the spectra exhibit a decrease in the integrated transfer over the probed energy window with increasing doping of the band insulator (Fig.~\ref{fig:VersusDoping}d). We compare the integrated transfer over the experimentally probed energy range at the $X$ point with the same quantity calculated from the theoretical spectra and find good agreement. This supports the interpretation that with increasing doping the peak is broadened and spectral weight is spread over a large range of energies due to high-energy particle-hole excitations of the Fermi sea that dress the magnon, as can be seen from the theoretical spectra (Fig.~\ref{fig:VersusDoping}e). We emphasize that this is a distinct mechanism from the trivial reduction in total transfer linear in doping that results from the lower starting atom number, as indicated by the dashed lines in Fig.~\ref{fig:VersusDoping}d. For this analysis, we focus on the $X$ point, where the energy corresponding to peak transfer remains approximately constant with doping, and the integration allows us to detect the small signal associated with the spectral weight redistribution.

A final property of interest in characterizing the polaron is its effective mass. We probe the polaron's dispersion by measuring photoinjection spectra at multiple momenta. For this measurement, we scan the injected momentum $\Delta \mathbf{k}$ along a line in the first BZ from the $M$ point towards the $\Gamma$ point 
(Fig.~\ref{fig:vsMomentum}a,b), and record the spectrum at each $\Delta \mathbf{k}$. For these scans, we analyze the centermost radial bin (average doping 8(1)\%) such that the magnon-polaron has  strong magnon character with a clear resonance. We repeat this process for a range of interactions, and show the resulting line centers in Fig. \ref{fig:vsMomentum}c. In this regime, we find that at all measured interactions the dispersions are well approximated by that of a single magnon, $E(k)=A(\mathrm{cos}(ka)-1)$, with a bandwidth $2A$ which is extracted as a fit parameter. From the bandwidth, the effective mass at the $M$ point may be calculated as the inverse curvature of $E(k)$. We display the calculated value as an inset in Fig.~\ref{fig:vsMomentum}c, and find good agreement with calculations (solid line).

In this work, we explored the evolution of magnons in a spin-polarized band insulator as hole dopants are introduced into our Fermi-Hubbard system. We find that the presence of mobile holes renormalizes the character of an injected magnon, resulting in the formation of a magnon-polaron, a collective spin excitation dressed by scattered dopants. This manifests in a shift in the quasiparticle energy and a redistribution of spectral weight. Our control over the injection momentum allows us to map out the dispersion of the quasiparticle and extract its effective mass. We use our experimental measurements of the dynamical spin structure factor to benchmark various state-of-the-art variational techniques for studying polarons in and out of equilibrium.

Looking forward, it would be interesting to extend our work to study magnon renormalization in other many-body states, including the doped Fermi-Hubbard antiferromagnet. More generally, we anticipate that Raman spectroscopy of Hubbard systems, the technique demonstrated in this work, will be a fruitful approach to studying other phenomena involving collective spin excitations in optical lattices. As a cold atom counterpart to inelastic neutron scattering in the solid state, it can be used to study a similarly wide range of physics in correlated systems~\cite{lovesey1984theory}. For example, in geometrically frustrated Hubbard systems, Raman spectroscopy may be be used to diagnose quantum spin liquid states by measuring excitation continua associated with the fractionalization of injected magnons into two spinons. The technique has also been proposed~\cite{EugeneIvanSpecTheory} as a way to measure the binding energy and dispersion of itinerant spin polarons recently observed in kinetically frustrated Hubbard systems~\cite{BakrDirectPolaronImage,GreinerDirectPolaronImage}. Finally, an exciting prospect would be to combine Raman spectroscopy with site- and spin-resolved imaging for studying the dynamical formation of these various types of polarons, starting from magnons injected with controlled momenta.

\textbf{Acknowledgements:} We thank Lawrence Cheuk for helpful discussions, and Areeq Hasan for assistance in early stages of this work. The experimental work was supported by the NSF (grant no. 2110475), the David and Lucile Packard Foundation (grant no. 2016-65128) and the ONR (grant no. N00014-21-1-2646). M.L.P. was supported by the NSF Graduate Research Fellowship Program.  D.A.H. was supported in part by NSF QLCI grant OMA-2120757. I.M.N. and E.D. were supported by the SNSF (project 200021\_212899), the Swiss State Secretariat for Education, Research and Innovation (contract number UeM019-1) and the ARO (grant no. W911NF-20-1-0163).
\\
\\
\noindent $^\dag$ Email: wbakr@princeton.edu



\bibliography{Polaron} 

\clearpage

\setcounter{figure}{0}
\setcounter{equation}{0}
\setcounter{section}{0}

\renewcommand{\thefigure}{S\arabic{figure}}
\renewcommand{\theHfigure}{S.\thefigure}
\renewcommand{\theequation}{S\arabic{equation}}
\renewcommand{\thetable}{S\arabic{table}}

\appendix

\begin{center}
    \large{\textbf{Methods}}
\end{center}
\vspace{-5mm}
\section{Experimental Setup}
Prior to spin-polarization, the preparation of a spin-balanced degenerate Fermi gas is similar to the scheme in our previous work \cite{brown2017spin}. Immediately after producing a degenerate spin-balanced gas consisting of states $\ket{1}$ and $\ket{3}$, we ramp the magnetic field to the zero-crossing of the 1-3 scattering length, and apply a resonant blowing pulse to remove the atoms in spin state $\ket{3}$ to prepare a spin-polarized gas in state $\ket{1}$. The pulse is 40 $\mu$s in duration at an intensity of $\sim0.06 \, I_{\mathrm{sat}}$ resonant with state $\ket{3}$, which is a nearly cycling transition at high field. This results in complete removal of atoms in $\ket{3}$, with 98.7(1)\% of atoms in $\ket{1}$ surviving. 

Once the atoms have been spin-polarized, the optical lattice is ramped to its final depth over 140 ms. For a fixed entropy per particle, the central filling and final temperature are determined by the strength of the confining potential relative to the tunneling rate $t$, with both increasing for higher confinements. The lattice depth of 10.6(1)$E_R$ used here was chosen to obtain a high central filling while keeping the tunneling rate reasonably large. All confinement in the lattice plane is provided by the waists of the lattice laser beams.

\begin{figure}[ht]
    \centering
    \includegraphics[width=\linewidth]{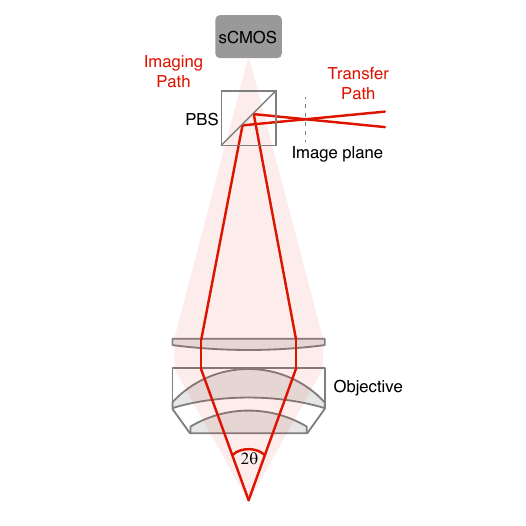}
    \caption{Experimental Schematic. Linearly polarized Raman beams (solid red lines) are reflected by a polarizing beam splitter (PBS) followed by half- and quarter-waveplates (not shown) for full polarization control to optimize the two-photon Rabi rate $\Omega_R$. Florescence imaging light (shaded area) is largely unpolarized and the $\sim 50\%$ that is transmitted through the PBS is collected on a CMOS camera used for single-atom imaging.}
    \label{fig:ramanSetup}
\end{figure}

With a spin-polarized band insulator prepared, we illuminate the sample with two Raman beams close to two-photon resonance with the $\ket{1}-\ket{2}$ transition at high field. Both beams are directed through the objective, as shown in Fig. \ref{fig:ramanSetup}. In this arrangement, the upper limit of the magnitude of the transferred momentum is $k_{\rm{max}}=k_0\,(2\cdot\rm{NA})$, where NA is the numerical aperture of the objective and $k_0 = 2\pi/\lambda_0$ is the wavevector of the Raman excitation light ($\lambda_0=671$~nm). In our case, NA = 0.5 is more than sufficient to cover the entirety of the first Brillouin zone of the optical lattice. 

The power in each beam is typically 25$~\mu$W with a waist of $\sim 80~\mu$m at the atom plane. Given that the analysis region for data versus doping extends out to a radius of 16 sites, we expect a maximal variation in the two-photon Rabi frequency for these data on the order of 5\%. The pulse time for most of spectra is set to 8~ms, such that the Fourier limit was well below the single-particle nearest-neighbor tunneling $t$. For the much broader single particle spectra at $\Delta\mathbf{k} \approx (0,\pi) a^{-1}$ and $(\pi,\pi)a^{-1}$presented in Fig.~\ref{fig:Figure1}e, pulse times of 4 ms and 2 ms is used, respectively. For the latter pulses the Raman optical power is increased so as to keep the total pulse area constant.

\section{Raman Coherence}

In this work, Raman transitions are driven between states in the hyperfine subspace $\Sigma \equiv \{ \onestate, \twostate \}$ via the intermediate $2P_{1/2}$ and $2P_{3/2}$ excited states. However, off-resonant excitation to the excited states can result in scattering out of $\Sigma$ into other hyperfine levels as well as extra decoherence within the science subspace. In heavier alkali atoms, the large fine structure splitting in the $P$ state enables Raman transfer with relative ease. However, the small splitting in Lithium results in fundamental limitations to the coherence of Raman transfer, and as such parameters must be finely optimized to enable our technique.

The effective Rabi frequency of transitions between states in $\Sigma$ can be found via the coherent sum of all pathways between excited states $\ket{e} \in \{2P_{1/2},2P_{3/2}\}$:
\begin{equation}
    \Omega_R = \sum_e \frac{\Omega_{1e}^1 \Omega_{2e}^2}{2 \Delta_e}
\end{equation}
where $\Omega_{ge}^\alpha \equiv -E_0^\alpha \matrixel{g}{\mathbf{d}\cdot \hat{\epsilon}^\alpha}{e}/\hbar$ is the single photon Rabi frequency between ground state $\ket{g}$ and excited state $\ket{e}$ from Raman beam $\alpha$ and $\Delta_e = \omega - \omega_e$ is the single photon detuning from $\ket{e}$. The optical intensity of a single Raman beam is related to the field by $I^\alpha = \frac{1}{2} \epsilon_0 c \abs{E_0^\alpha}^2$. 

For a single atom the total rate of scattering from ground state $\ket{g}$ induced by the Raman beams is calculated as
\begin{equation}
    \Gamma_R^g = \sum_e \sum_\alpha \Gamma \frac{ \abs{E^\alpha_0 \matrixel{g}{\mathbf{d}\cdot\hat{\epsilon}^\alpha}{e}}^2}{4\hbar^2 \Delta_e^2},
\end{equation}
where $\Gamma\approx2\pi\times6$MHz is the decay rate of the relevant excited states and the total scattering rate is $\Gamma = \sum_g P_g \Gamma^g_R$. This total scattering rate can be resolved into two components: state-changing Raman (inelastic) and state-preserving Rayleigh (elastic) scattering. For applications where information is encoded solely in internal degrees of freedom, Rayleigh scattering events do not lead to depolarization or decoherence. However, given that the scattered photon does impart recoil momentum to the atom, motional state purity will be affected. For our scheme, this results in a distinguishable transition out of the coupled state pair $\ket{\uparrow,n,\mathbf{q}} \leftrightarrow \ket{\downarrow,n,\mathbf{q}+\Delta \mathbf{k}}$, and therefore all scattering events are considered decohering.

Both Raman beams also contribute light shifts to the ground states according to:
\begin{equation}
    \hbar \delta_{LS}^g = \sum_e \sum_\alpha \frac{ \abs{E^\alpha_0 \matrixel{g}{\mathbf{d}\cdot\hat{\epsilon}^\alpha}{e}}^2}{4\hbar \Delta_e}
\end{equation}

\begin{figure}[t]
    \centering
    \includegraphics[width=\linewidth]{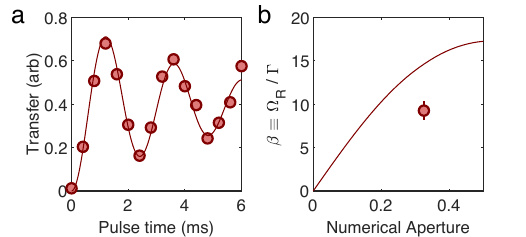}
    \caption{Raman coherence. (a)Measured Rabi oscillations between states $\ket{1}$ and $\ket{2}$ driven with co-propagating Raman beams (marker) with fit to a damped Rabi oscillation (curve), yielding $\beta=9.3(1.0)$. Beams propagate at an angle of $19^\circ$ with respect to the quantization axis. (b) Theoretically calculated maximal coherence $\beta$ optimized over all polarizations versus numerical aperture of the co-propagating beams (curve). Experimental measurement from (a) at NA$=\sin(19^\circ)$ is displayed for comparison. Measurements are done in a uniform magnetic field of $591$ Gauss, with a single photon detuning $\sim 4$ GHz to the red of the $D_1$ line at zero field.}
    \label{fig:ramancoherence}
\end{figure}

We now consider computation of the above matrix elements. While standard matrix elements may be used in the $\ket{LSIJFm_F}$ basis, the combination of the small fine structure of $^6\mathrm{Li}$ and the high fields used in this work result that perturbations to $L-S$ coupling are important. For this reason, we use the fully uncoupled basis $\ket{LSIm_Lm_Sm_I}$ for all calculations.

Neglecting the small nuclear electric quadrupole moment, there are 7 contributions to the internal state of the atom \cite{lyons1970theoretical}:
\begin{align}
    H_{\mathrm{int}} = \mathbf{B}\cdot \mu_B (g_L \mathbf{L} + g_S \mathbf{S} + g_I \mathbf{I}) + c_{\mathrm{fs}} \mathbf{L}\cdot\mathbf{S} \\+ c_{\rm{orb}} \mathbf{L}\cdot\mathbf{I} + c_{\rm{cont}} \mathbf{S}\cdot\mathbf{I} + c_{\rm{dip}} \mathbf{I}\cdot(\mathbf{S}\otimes\mathbf{C}^{(2)})^{(1)}
\end{align}
In order these are the Zeeman, fine structure, orbital, Fermi contact and dipolar terms. The latter three are responsible for hyperfine coupling. While J is not strictly a good quantum number, for $c_{\rm{fs}} \gg c_{\rm{orb}},c_{\rm{cont}},c_{\rm{dip}}$ we can project the above terms onto the more familiar approximate hyperfine interaction $A_J \mathbf{I}\cdot \mathbf{J}$ in order to evaluate the terms. Many-body electron calculations $^6\mathrm{Li}$ and $^7\mathrm{Li}$ have indicated that contributions to the effective hyperfine interaction are dominated by the contact and orbital terms \cite{lyons1969many}, however, for simplicity we choose to neglect the more complicated orbital term.

We compute the marix elements to be \cite{edmonds1996angular,levy1965symmetrical}:
\begin{align}
    \matrixel{JFm}{\mathbf{S}\cdot\mathbf{I}}{J'Fm} = (-1)^{\Sigma + J+J' + F + 1} \sqrt{\chi(I) \chi(S)} \\\left((2J + 1) (2 J' + 1) \right)^{1/2}
    \Gj{F}{I}{J}{1}{J'}{I} \Gj{L}{S}{J}{1}{J'}{S},
\end{align}
where $\chi(L) \equiv (2L + 1)(L+1)L$.
The diagonal elements represent the projection onto the $\mathbf{I}\cdot\mathbf{J}$ Hamiltonian and can be shown to be equivalent to the well-known projection theorem $\expval{\mathbf{L}\cdot\mathbf{I}}_{JFm} = (\mathbf{L}\cdot\mathbf{J})(\mathbf{I}\cdot\mathbf{J})/\mathbf{J}^2$. Following a similar procedure for the contact term, we obtain
\begin{align}
    c_{\rm{orb}} =& 1/2(A_{3/2} + A_{1/2})\\
    c_{\rm{cont}} =& 2A_{3/2} - A_{1/2}
\end{align}
From this we diagonalize the above Hamiltonian numerically for all fields to identify the composition of the excited states. We verify the fall off in Raman coupling with magnetic field \cite{wei2013magnetic} as the field becomes strong with respect to fine and hyperfine structure.

Yet another restriction on coherence is the orientation of the beams with respect to the quantization axis, which is aligned to the optical axis of the microscope objective through which the Raman beams pass. As the transition $\ket{1}$ to $\ket{2}$ must involve a $\pi$ polarization component, coherent Raman coupling will go to zero as the pointing of the Raman beams is tuned closer to parallel with the quantization axis. To probe the scattering-limited coherence in our system we align the Raman beams such that they are co-propagating (with common polarization) at an angle $\theta$ with respect to the quantization axis and measure Rabi oscillations between $\ket{1}$ and $\ket{2}$ (Fig. \ref{fig:ramancoherence}a). From this measurement we extract $\beta=\Omega_R / \Gamma$, or the ratio of the coherent Raman coupling to the total scattering rate. We compare this to the theoretically calculated maximum $\beta$ versus numerical aperture $\mathrm{NA}=\sin(\theta)$, and find that we achieve approximately $65\%$ of the theoretical maximally achievable $\beta$ (Fig. \ref{fig:ramancoherence}b). We expect that this reduction can largely be explained by imbalanced intensities at the atoms, additional transverse decoherence (as would be induced by fluctuating light shifts and magnetic fields) or imperfect spectral filtering of the light we use to drive the transition.

\section{Spatial matrix elements}
\begin{figure}[t]
    \centering
    \includegraphics[width=\linewidth]{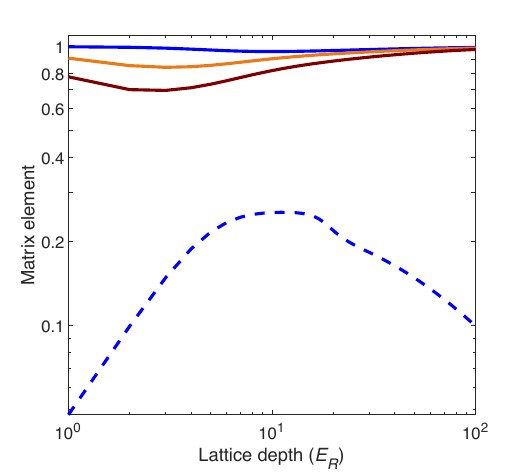}
    \caption{Spatial matrix elements in the optical lattice. Plotted is the magnitude of the matrix element of the momentum kick operator $e^{i \Delta \mathbf{k} \cdot \hat{\mathbf{x}}}$ between the ground Bloch eigenstate for our lattice potential $\ket{n =0,\mathbf{q} = 0}$ and the relevant excited state $\ket{n',\mathbf{q}'}$, i.e. $|\matrixel{n',\mathbf{q}'}{e^{i \Delta \mathbf{k} \cdot \hat{\mathbf{x}}}}{0,0}|$ for various values of $n'$ and $\mathbf{q}'$. The upper curves represent transitions within the ground band ($n' = 0$) and have $\Delta \mathbf{k} = \mathbf{q}' = (0.5,0.5) \pi/a$ (blue), $(0.95,0)\pi/a$ (orange) and $(0.95,0.95)\pi/a$ (brown). The lower (dashed) curve by contrast has $n'=1$ and $\Delta \mathbf{k} = \mathbf{q}'=(0.5,0.5)\pi/a$ and shows the significant suppression of band-changing transitions in the lattice.}
    \label{fig:matrixElements}
\end{figure}

Owing to the non-zero momentum imparted by the Raman beams, overlap with the final external state must also be taken into account when computing the resonant Rabi frequency as
\begin{align}
    \matrixel{\psi_f}{H}{\psi_i} =& \matrixel{\psi_f}{\frac{\hbar\Omega_R}{2} (\sigma^+ e^{i\,  \Delta \mathbf{k} \cdot \mathbf{r}} + \mathrm{h.c.})}{\psi_i}\\ 
    =& \frac{\hbar \Omega_R}{2} \matrixel{i}{\sigma^+}{f} \matrixel{\psi_f(\mathbf{r})}{e^{i \, \Delta\mathbf{k}\cdot \mathbf{r}}}{\psi_i(\mathbf{r})},
\end{align}
where $\ket{i}$ denotes the internal (spin) state and $\ket{\psi_i}$ denotes the spatial wavefunction. The atoms experience a tightly confining vertical lattice in the $z$ direction and a tunnel coupled lattice in the $x-y$ plane. Therefore, the full spatial wavefunction is separable into the product of a ground state harmonic oscillator in $z$ and Bloch functions in the 2D plane, i.e.
\begin{align}
    \psi_i(\mathbf{r}) = \psi^{n = 0}_{\mathrm{SHO}}(z) \psi_{n,\mathrm{\mathbf{q}}}(x,y),
\end{align}
where $\psi_{n,\mathbf{q}}(x,y)$ is the Bloch eigenvector of band $n$ at quasimomentum $\mathbf{q}$. The matrix elements of the momentum displacement operator can then be computed using the decomposition of the eigenvectors into Bloch functions. Consider first the case of the eigenfunctions of a 1D lattice $\ket{n,q}$ with band index $n$ and quasimomentum $q$:
\begin{gather}
    \matrixel{n',q'}{e^{i \Delta k \hat{x}}}{n, q} = \int dx \; \psi_{n',q'}^* (x) e^{i \Delta k x } \psi_{n,q}(x)\\
     = \int dx \; u_{n',q'}^*(x) e^{i \Delta k x} e^{i (q - q') x} u_{n,q}(x) \\
      = \sum_{k',k} u^*_{n',q'}(k') u_{n,q} (k) \delta((k + (\Delta k - \Delta q)) - k').
\end{gather}
In the above, $k, k'\in \mathbb{Z} (2 \pi/a)$ are elements of the reciprocal lattice while $\Delta k$ and $ \Delta q\equiv q'-q$ are arbitrary momenta. Thus unless $\mathrm{mod}(\Delta k - \Delta q,k_r) = 0$ the matrix element is zero, which reflects conservation of momentum up to a reciprocal lattice vector $2\pi/a$. In the case of driving a transition where $\Delta k $ from the Raman beams lies within the first Brillouin zone, this matrix element becomes
\begin{align}
    \matrixel{n',q'}{e^{i \Delta k \hat{x}}}{n, q} = \delta(\Delta k -\Delta q) \bra{u_{n',q'}}\ket{u_{n,q}}.
\end{align}


The above expression can be easily generalized to higher dimensions, and in Fig.\,\ref{fig:matrixElements} we plot the matrix elements $\matrixel{n',\mathbf{q}'}{e^{i \mathbf{\Delta k} \cdot \mathbf{r}}}{n,\mathbf{q}}$ for the 2D retro-reflected lattice that is used in this paper at several momenta representative of the range explored in this work. From this it can be seen that matrix elements for band-preserving transitions within the first Brillouin zone are only modestly suppressed at any lattice depth. Intra-band transitions on the other hand are suppressed at all lattice depths, in addition to being far detuned by the lattice band gap.

\section{Raman Momentum Calibration}

Here we detail the calibration method of the momentum imparted to the atoms during Raman spectroscopy. This measurement is taken in a reasonably deep $33 E_R$ lattice such that tunneling is suppressed and atoms remain predominantly in the motional ground state of individual lattice wells ($\eta_{x,y}^2 \equiv \hbar \Delta k^2_{x,y}/(2 m \omega_{latt})\approx 0.1$) while lattice light-shift induced decoherence that would reduce the signal contrast is minimized. If the Raman beams are pulsed on for a variable duration of time, the spatially varying phase which is present in the drive will imprint onto the atoms in the equatorial plane. A measurement in the $z$ basis will therefore reveal no information about the differential phase accrued across the cloud, thus a global RF pulse is used to rotate the measurement basis to the $x-y$ plane. The spatially dependent internal state of the atom after the pulse sequence is
\begin{align}
    \ket{\psi_f(\mathbf{r})}= \hat{U}_{\mathrm{RF}} \,\hat{U}_{\mathrm{R}}(\mathbf{r}) \ket{\downarrow},
\end{align}
where $\hat{U}_{\mathrm{RF}} \equiv \exp \small(-it\hat{H}_{\mathrm{RF}}\small)$ and $\hat{U}_{\mathrm{R}}(\mathbf{r}) \equiv \exp\small({-it_{R}\hat{H}_{\mathrm{R}}(\mathbf{r})}\small)$. The Hamiltonians are:
\begin{align}
    H_{\mathrm{R}}&=\frac{\hbar \Omega_R}{2}(e^{i\mathbf{k}\cdot\mathbf{r}} \hat{\sigma}^+ + c.c.)+ \frac{\hbar\Delta}{2} \hat{\sigma}_z,\\
H_{\mathrm{RF}}&=\frac{\hbar \Omega}{2}( e^{i\phi}\hat{\sigma}^+ + c.c.),
\end{align}
where in the above expressions we have accounted for a light shift $\Delta$ due to the Raman beams and a phase $\phi$ between the RF and Raman transitions. We have also demoted $\hat{\mathbf{r}}$ from an operator to a scalar as atoms are confined to a single lattice site. For the simplifying case of resonant excitation $\Delta = 0$ and RF pi/2 pulse $\Omega t = \pi/2$, the final density of $\ket{\uparrow}$ atoms in the lattice $n_\uparrow(\mathbf{r})$ can be calculated from the excitation probability $\abs{\braket{\uparrow}{\psi_f(\mathbf{r})}}^2$ which yields
\begin{equation}
    n_\uparrow(\mathbf{r})= n_0 \frac{1}{2} (\cos (\mathbf{k}\cdot\mathbf{r}-\phi ) \sin \Omega_R t_R +1),
\end{equation}
where $n_0$ is the initial density of $\ket{\uparrow}$ atoms in the lattice. Therefore, in a given fluoresence image of the experiment the probability of excitation will be modulated at the spatial frequency of $\mathbf{k}$. Stochastic fluctuations in the RF phase $\phi$ shift the observed density pattern randomly, causing the experimentally averaged density signal to display no modulation, $\langle n_\uparrow(\mathbf{r})\rangle_\phi = n_0/2$. Instead  we evaluate the two-point correlation function as
\begin{align}
    C^{(2)}(\mathbf{r},\mathbf{r}+\boldsymbol{\delta}) &\equiv \langle n_\uparrow(\mathbf{r}) n_\uparrow(\mathbf{r}+\boldsymbol{\delta}) \rangle_\phi - \langle n_\uparrow(\mathbf{r})\rangle_\phi \langle n_\uparrow(\mathbf{r} + \boldsymbol{\delta})\rangle_\phi \\
    &= \frac{n_0^2}{8} \cos (\mathbf{k}\cdot\boldsymbol{\delta})\sin ^2\Omega_R t_R,
\end{align}
which also has the effect of removing the dependence on the chosen starting position $\mathbf{r}$ of the correlator.

\begin{figure}[t]
    \centering
    \includegraphics[width=\linewidth]{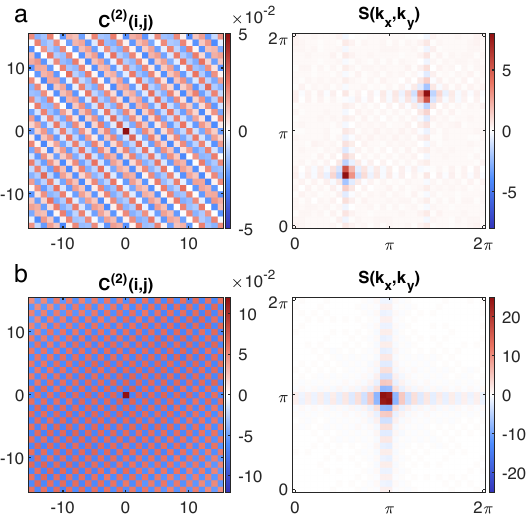}
    \caption{Raman momentum transfer measurement.  Real-space correlations $C^{(2)}(\bf{r})$ and accompanying static spin structure factor $\tilde{S}(\bf{k})$ shown for $\Delta \mathbf{k} =  (0.6057(5),0.5978(5)) \, \pi /a $ (a) and $\mathbf{k} = (1.0093(4),0.9940(4)) \, \pi/a$ (b). For these measurements different pulse times are used ($t_R = 900 \mu$s and $600 \mu$s, respectively) owing to the reduction in coherent coupling as $\Delta \mathbf{k}\to \mathbf{0}$.}
    \label{fig:correlationsMeasuremet}
\end{figure}

We also note that in the presence of finite detuning the expression for $n_{\uparrow}(\mathbf{r})$ is more complicated, but still only shows density modulation at $\mathbf{k}$, albeit with an additional phase dependent on the driving parameters, i.e.
\begin{align}
    n_\uparrow(\mathbf{r}) = \frac{n_0}{2 \tilde{\Omega}_R^2}\bigg[ \tilde{\Omega}_R^2&+  \Omega_R \tilde{A}_R   \cos\small(\mathbf{k}\cdot\mathbf{r} - \phi -\tilde{\phi}_R\small)\sin  \Omega t  \\&- \left(\Delta ^2+\Omega_R ^2 \cos \tilde{\Omega}_R t_R \right) \cos \Omega t \bigg],
\end{align}
where we have introduced the quantities $\tilde{A}_R$ and $\tilde{\phi}_R$ as
\begin{gather}
    \tilde{A}_R^2 = 2 \sin ^2(\tilde{\Omega}_R t_R/2) (2 \Delta ^2+\Omega_R ^2 \cos \tilde{\Omega}_R t_R+\Omega_R ^2)\\
    \tan\tilde{\phi}_R = \frac{\Delta  \tan \small(\tilde{\Omega}_R t_R/2 \small)}{\tilde{\Omega}_R},
\end{gather}
and $\tilde{\Omega}_R = \sqrt{\Omega_R^2 + \Delta^2}$ is the generalized Rabi frequency of the Raman transition. In Fig \ref{fig:correlationsMeasuremet}, we show a sample of real-space correlations and the accompanying spin structure factor for multiple values of $\Delta \mathbf{k}$.

\section{Quasiparticle Lifetime and Spectral Width}
\begin{figure}[t]
    \centering
    \includegraphics[width=\linewidth]{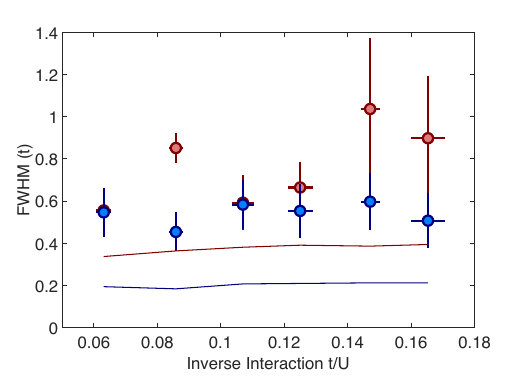}
    \caption{Magnon-polaron linewidth vs interactions. Experimentally measured spectral linewidths (markers) are reported as the half-width at half-max (HWHM) of a Lorentzian fit to the data versus inverse interaction $t/U$ for injected momentum $\Delta \mathbf{k} \approx (\pi ,\pi) a^{-1}$ (red markers) and $\Delta \mathbf{k} \approx (0,\pi)a^{-1}$ (blue markers). For consistency with experimental data analysis, theory curves are also fit to a Lorentzian with the HWHM plotted (solid lines).}
    \label{fig:interactionLinewidth}
\end{figure}

For quasiparticle excitations, the lifetime is calculated as the inverse width of the spectral function resonance \cite{schmidt2011excitation}, which is in principle directly measurable from the recorded spectra. However, sources of technical broadening beyond that introduced by many-body effects (such as doping and interactions) result in an inaccurate estimate of the quasiparticle lifetime. Technical sources of broadening in our setup consist of inhomogenous broadening, finite temperature and the external trapping potential. We find that single-particle inhomogenous broadening is negligible at our chosen pulse time of 8 ms as this is less than the bare coherence time $T_2^*\approx35$ ms of our system. As such, we verify that the fitted widths of the deep lattice scans (such as in Fig. \ref{fig:figure2}a) are consistent with the Fourier limit. 

Another potential source of broadening is the temperature of the initial spin-polarized gas in the lattice. From the analysis of nearest-neighbor density-density correlations, we find a temperature of $T/t=1.9(3)$, averaged over experiments versus interactions, doping and momentum. We compare the experimentally measured linewidths to those obtained from theoretical spectra calculated at finite temperature in Fig. \ref{fig:interactionLinewidth} and find that the theoretically calculated values systematically under-predict the measured linewidths, indicating an additional source of broadening that isn't present in the finite temperature model. 

We note that the increased temperature in this work is directly related to working with a spin-polarized sample at high filling. For a two-component gas, increasing interactions results in a higher specific heat capacity as low-energy spin excitations carry the entropy \cite{EhsanEntropyPhysRevA.84.053611,cocchi2017measuring}. Additionally, the low compressibility of the desired final state requires strong confinement, which, unless paired with significant entropy re-distribution, results in higher overall temperatures. We estimate the average entropy per particle in the initial state of around $10^3$ atoms in the lattice to be 0.9(2)$k_B$, which is a typical value achieved in similar experiments after evaporation \cite{TroyerEntropyPhysRevLett.104.180401,BlumerEntropyPhysRevA.85.061602,GrossEntropyPhysRevLett.115.263001}.

\begin{figure}[t]
    \centering
    \includegraphics[width=\linewidth]{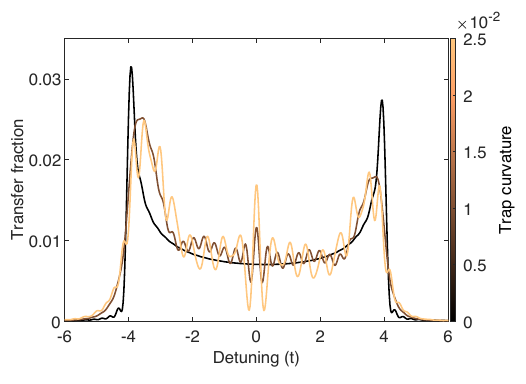}
    \caption{Simulation of non-interacting fermions showing the result of increasing trap curvature. This simulation has the parameters $T/t = 2$, $n_{\mathrm{max}}=0.93$, with a pulse time $t_r = 8 \pi (\hbar/t)$ and Rabi frequency $\hbar\Omega_r = t/20$. The pulse time here is increased beyond that of the main text Fig. \ref{fig:Figure1} to isolate trap-induced broadening. Trap curvature refers to the dimensionless parameter $F \equiv m\omega^2a^2/2t$ where $\omega$ is the trapping frequency of the harmonic confinement in the lattice.}
    \label{fig:niTrapBroadening}
\end{figure}

We find the most likely source of the observed broadening is the external trapping potential present in the lattice. While this is required to maintain a high central filling, the non-uniform potential breaks the lattice translational symmetry, meaning that quasimomentum is no longer conserved and the Raman pulse may excite a range of final states, and therefore energies. We quantify the resulting trap-induced broadening in the non-interacting case with a numerical calculation including the effects of finite temperature, filling and trap curvature in Fig. \ref{fig:niTrapBroadening}. This simulation is performed in 1D such that a large lattice can be used to ensure convergence, but similar conclusions can be drawn for the full 2D case, which is used for comparisons with data in main text Fig. \ref{fig:Figure1}. 

We consider the initial density matrix $\rho$ to be that of a thermal state of the 1-D tight binding Hamiltonian $H_{\mathrm{tb}} = -t\sum_{\langle i,j\rangle}(\hat{c}_{i,\sigma}^{\dagger}\hat{c}_{j,\sigma}+\mathrm{h.c})+\frac{1}{2} m \omega^2 a^2 \sum_i (i - i_0)^2$ with harmonic trapping of frequency $\omega$ and nearest-neighbor tunneling $t$. The global chemical potential $\mu$ is chosen so as to reproduce the observed maximum filling in the center of the lattice. We then time-evolve the initial state under the total Hamiltonian $H = H_{\mathrm{tb}} + H_{R}$ and plot the transfer versus detuning for various values of $\omega$, which can be quantified in terms of the dimensionless parameter $F \equiv m \omega^2 a^2/2t$ (Fig. \ref{fig:niTrapBroadening}). A long pulse time is chosen to reduce Fourier broadening. Interestingly, we find that temperature plays no role in broadening the resulting spectra, while increasing harmonic confinement leads to significant redistribution of spectral weight around the originally sharp peaks at $\Delta = \pm 4 t$. We also find that the reduction in maximum filling does not lead to excess broadening, but does explain the asymmetry in the peaks observed in main text Fig. \ref{fig:Figure1}. 

The LLP transformation utilized here is incompatible with the breaking of translational symmetry in the lattice and this effect cannot be added easily to our many-body simulations. However, the broadening mechanism is fundamental and we expect that most of the extra broadening we see in the spectra are attributable to the non-uniform lattice potential used here.

\section{Theoretical framework}
In this section, we present the theoretical framework used to capture the experimental results (further details in~\cite{morera2025upcoming}). Upon injecting the spin-$\downarrow$ using the Raman protocol, entanglement develops between the injected fermion and the majority spin-$\uparrow$ fermions.
In order to capture this entanglement, we perform a unitary Lee-Low-Pines (LLP) transformation which generates
a coherent superposition of the spin-$\downarrow$ fermion at different locations in a correlated way with the many-body wavefunction of the majority of fermions. The LLP transformation is given by
\begin{align}
    \hat{U}_{\rm LLP} = e^{i \hat{\mathbf{r}}_{\downarrow} \hat{\mathbf{Q}}_{h}},
    \label{Eq:LLP}
\end{align}
where $\hat{\mathbf{r}}_{\downarrow} = \sum_i \mathbf{r}_i \hat{c}^{\dagger}_{i,\downarrow}\hat{c}_{i,\downarrow}$ is the position of the spin-$\downarrow$ fermion and $\hat{\mathbf{Q}}_{h} = \sum_{\mathbf{q}} \mathbf{q} \hat{h}^{\dagger}_{\mathbf{q}}\hat{h}_{\mathbf{q}}$ is the total momentum of the doped holes, with $\hat{c}^{\dagger}_{\mathbf{q},\uparrow}=\hat{h}_{\mathbf{q}}$. 
The LLP transformation can also be understood as a change of reference frame from the lab to the frame of the spin-$\downarrow$ fermion. 
To perform calculations, we need to obtain the operators in the new reference frame. For example, the Fermi-Hubbard Hamiltonian in the LLP frame is given by
\begin{align}
    \hat{H}_{\rm LLP}({\mathbf{K}})&\equiv \hat{U}_{\textrm{LLP}}^{\dagger}\hat{H}\hat{U}_{\textrm{LLP}} = t_\uparrow \sum_{\mathbf{k}, \boldsymbol{\delta}} e^{-i \mathbf{k} \cdot \boldsymbol{\delta}} h_{\mathbf{k}}^{\dagger}h_{\mathbf{k}}  + U\nonumber \\
    &- t_\downarrow\sum_{\boldsymbol{\delta}} e^{-i \left(\mathbf{K}-\hat{Q}_{h} \right) \cdot \boldsymbol{\delta}}
     - \frac{U}{N_s} \sum_{\mathbf{k},\mathbf{q}}\hat{h}^{\dagger}_{\mathbf{k}}\hat{h}_{\mathbf{q}}.
\end{align}
The spin-$\downarrow$ fermion is eliminated from the Hamiltonian in the new frame, but additional interactions between the spin-$\uparrow$ fermions appear in the new frame. 

Given the entanglement generated by the LLP transformation, we proceed to approximate the many-body wavefunction of the majority of spin-$\uparrow$ fermions as a Gaussian state. In order to obtain the ground state, we perform an imaginary time evolution in the projected Gaussian manifold with a typical imaginary time step $d\tau=10^{-2}$ and with an energy error $\delta E = 10^{-4}$. 

The finite temperature Raman photoexcitation spectrum can be obtained by performing a regularized Fourier transform of the time dependent spin structure factor,
\begin{equation}
    S_\beta^-(\mathbf{k},t) = \textrm{Tr}\{\hat{S}^+_{\mathbf{k}}(t) \hat{S}^-_{\mathbf{k}} \hat{\rho}_{\beta} \}.
    \label{Eq:Ot}
\end{equation}
To conduct the real time evolution we employ the Dirac's variational principle with a time step $dt=5 \times10^{-3}$. Finite temperature effects can be taken into account by Monte Carlo sampling different time evolutions following the density matrix $\hat{\rho}_{\beta}$, which we consider to be a thermal Boltzmann distribution of Slater determinants, as the initial state is solely composed of spin-$\uparrow$ fermions. The energies in Fig.~\ref{fig:VersusDoping}b of the main text are extracted by fitting a single Lorentzian to the theoretical spectra. This is the same approach used for the experimental spectra, which do not resolve the emergence of the weak second higher energy peak for large dopings seen in the theoretical spectra (Fig.~\ref{fig:VersusDoping}e). This could be due to experimental limitations, such as technical broadening and the measurement's signal-to-noise ratio, or alternatively, the peak might be an artifact of the approximate theoretical technique.

\section{Non-interacting photoexcitation}
In the non-interacting regime $U/t=0$, we can obtain an exact expression of the Raman photoexcitation. The dynamical
spin structure factor $S_\beta^-(\mathbf{k},\omega)$ can be related to the spin susceptibility $\chi_{\beta}(\mathbf{k},\omega)$
\begin{align}
    S^-_{\beta}(\mathbf{k},\omega) &= \frac{1}{\pi} \lim_{\eta\rightarrow 0}\textrm{Im}\chi_{\beta}(\mathbf{k},\omega), \\
    \chi_{\beta}(\mathbf{k},\omega) &= \sum_{\Phi,m} p(\Phi) \frac{|\langle m |\hat{S}^-_{\mathbf{k}}| \Phi(\{\mathbf{q}_i\}) \rangle|^2 }{\omega +i\eta - (E_m-E_{\Phi})}.
    \label{Eq:FDT}
\end{align}
In the non-interacting case $U/t=0$, the spin susceptibility can be simplified to,
\begin{align}
    \chi_{\beta}(\mathbf{k},\omega) =\frac{1}{N_s} \sum_{\Phi,\mathbf{q}} p(\Phi) \frac{1-n^{(\Phi)}_{\mathbf{q}} }{\omega +i\eta - E_{\mathbf{q},\mathbf{k}}},
\end{align}
with $E_{\mathbf{q},\mathbf{k}}=\epsilon_h(\mathbf{q}) + \epsilon_{\downarrow}(\mathbf{k}-\mathbf{q})$, $N_s$ being the number of sites, $\epsilon_h(\mathbf{q})=-\epsilon_\uparrow(\mathbf{q})$, $\epsilon(\mathbf{q})=-2t\sum_{\pmb{\delta}}\cos(\mathbf{q}\cdot \pmb{\delta})$ the single-particle energies, and $n^{(\Phi)}_{\mathbf{q}}$ the occupation number of momentum $\mathbf{q}$ for the Slater determinant $\ket{\Phi}$. We observe that the spin susceptibility $\chi_{\beta}(\mathbf{k},\omega)\equiv \chi^{(\rm BI)}(\mathbf{k},\omega)-\chi^{(\rm h)}_{\beta}(\mathbf{k},\omega)$ consists of two contributions: the first comes from the spin susceptibility of the fully polarized state, while the second arises from the thermally occupied holes. The former, can be easily computed,
\begin{align}
    \chi^{(\rm BI)}(\mathbf{k},\omega) = \frac{2}{\pi}\frac{1}{\sqrt{z^2-(a-b)^2}}K\left(\sqrt{\frac{4ab}{z^2-(a-b)^2}}\right),
\end{align}
where $z=w+i\eta$, $a=4t\sin(k_x/2)$, $b=4t\sin(k_y/2)$ and $K(x)$ is the complete elliptic integral of the first kind. The contribution of the band insulator to the spectral function coincides with the density of states (DOS) of the hole-magnon problem at total momentum $\mathbf{k}$. Interestingly, the spectral function along the path $\Gamma\rightarrow X$ ($\Gamma \rightarrow M$) reduces to the one-(two-)dimensional DOS with an effective hopping $2t\sin(k/2)$,
\begin{align}
    A^{(\rm BI)}(\mathbf{k},\omega) = 
    \begin{cases}
    \frac{\theta\left(a-|\omega|\right)}{\pi \sqrt{a^2-\omega^2}} & \rm for \quad \mathbf{k}=(k,0), \\
    \frac{1}{a\pi^2} K\left(\sqrt{1-\left(\frac{\omega}{2a}\right)^2}\right) & \rm for \quad \mathbf{k}=(k,k).
    \end{cases}
\end{align}
The spectral function contribution of the band insulator background contains a continuum of particle-hole excitations which is bounded by $4t (\sin(k_x)+\sin(k_y))$. Moreover, it shows a divergence at $\omega=0$ for $k_x=k_y$  and at the edge of the spectrum for $k_y=0$ or $k_x=0$.

\end{document}